\documentclass[a4paper,11pt]{article}
\usepackage{jheppub}

\usepackage[utf8]{inputenc}
\usepackage{xcolor}
\usepackage{color}
\usepackage{relsize}
\usepackage{graphics}
\usepackage{epstopdf}
\usepackage{hyperref}
\usepackage{mathrsfs}
\usepackage{ragged2e}
\usepackage{amssymb}
\usepackage{placeins}
\usepackage{comment}

\usepackage[normalem]{ ulem }
\usepackage{amsthm}
\usepackage{amsmath}
\usepackage{hyperref}
\usepackage{cancel}
\newcommand{\be}{\begin{equation}}
\newcommand{\ee}{\end{equation}}
\newcommand{\bea}{\begin{eqnarray}}
\newcommand{\eea}{\end{eqnarray}}

\newcommand{\vev}{ {\langle\phi\rangle} }

\newcommand{\QCD}{{\rm QCD}}

\def\ie{{\it i.e.},~}
\def\eg{{\it e.g.},~}

\newcommand{\beq}{\begin{equation}}
\newcommand{\eeq}{\end{equation}}
\newcommand{\beqn}{\begin{eqnarray}}
\newcommand{\eeqn}{\end{eqnarray}}
\newcommand{\nn}{\nonumber}
\newcommand{\LambdaQCD}{\Lambda_{\mathrm{QCD}}}
\newcommand{\LambdaQCDSM}{\Lambda_{\mathrm{QCD}}^{0}}
\newcommand{\LambdaEW}{T_{\mathrm{EW}}^{0}}

\newcommand{\Tosc}{T_{\mathrm{osc}}}
\newcommand{\Tup}{T_{\uparrow}}
\newcommand{\Tdown}{T_{\downarrow}}

\usepackage{pgfplots}

\title{Resurrecting Low-Mass Axion Dark Matter Via a Dynamical QCD Scale}
\preprint{UCI-HEP-TR-2021-13, IPPP/20/98}

\author[a]{Lucien Heurtier,}
\author[b,c]{Fei Huang}
\author[c]{and Tim M.P. Tait}

\affiliation{\vspace{15pt}$^a$ Institute for Particle Physics Phenomenology, Durham University, South Road, Durham, U.K.}
\affiliation{$^b$ CAS Key Laboratory of Theoretical Physics, Institute of Theoretical Physics,
Chinese Academy of Sciences, Beijing 100190, China
}
\affiliation{$^c$ Department of Physics and Astronomy, University of California, Irvine, CA 92697 USA}

\emailAdd{lucien.heurtier@durham.ac.uk}
\emailAdd{huangf4@uci.edu}
\emailAdd{ttait@uci.edu}

\abstract{
In the framework where the strong coupling is dynamical, the QCD sector may confine at a much higher temperature than it would in the Standard Model, and the temperature-dependent mass of the QCD axion evolves in a non-trivial way. 
We find that, depending on the evolution of $\Lambda_{\rm QCD}$, the axion field may undergo multiple distinct phases of damping and oscillation leading generically to a suppression of its relic abundance.  
Such a suppression could therefore open up a wide range of parameter space, resurrecting in particular axion dark-matter models with a large Peccei-Quinn scale $f_a\gg 10^{12}~\mathrm{GeV}$, \ie with a lighter mass than the standard QCD axion.}

\begin{document}
\maketitle

\section{Introduction}
The nature of the dark matter whose existence is required to explain astronomical and cosmological observations remains among the
most pressing questions confronting particle physics.  A large
suite of ongoing and planned 
experiments seeks to detect it and understand the role it
plays in a fundamental description of nature containing the Standard Model (SM) \cite{Bertone:2018krk}.

The axion stands out
among candidates to play the role of particle dark matter \cite{Preskill:1982cy,Abbott:1982af,Dine:1982ah} as one whose existence has been independently postulated to explain another mystery in particle physics: the apparent lack of violation of charge
conjugation-parity (CP) symmetry by quantum chromodynmaics (QCD), the strong nuclear interaction, as demonstrated
by the absence of an observable electric
dipole moment for the neutron \cite{Afach:2015sja}.  The axion $a(x)$ arises
as a pseudo Nambu-Goldstone boson \cite{Weinberg:1977ma,Wilczek:1977pj} associated with the spontaneous breaking of a Peccei-Quinn U(1)$_{\rm PQ}$ symmetry \cite{Peccei:1977ur,Peccei:1977hh} whose presence insures that CP will be conserved by the strong nuclear force.
This symmetry is anomalous with respect to
SU(3)$_{\rm C}$, and the explicit breaking by
strong instantons induces a periodic potential of the form $V_{\rm PQ}(a)\propto \Lambda_{\rm QCD}^4 \cos\left(a / f_a\right)$, resulting in a
mass for the axion of order $\Lambda_{\rm QCD}^2 / f_a$, where $f_a$ characterizes the scale at which the PQ symmetry is broken.  As a massive neutral particle with
feeble interactions with the Standard Model (SM)
\cite{Kim:1979if,Shifman:1979if,Dine:1981rt,Zhitnitsky:1980tq}, the axion has all of the correct properties necessary to play the role of dark matter.

The relic abundance of axions depends sensitively on the cosmological history of the early Universe (see Ref.~\cite{Marsh:2015xka} for a review).  If the PQ symmetry is broken after inflation, the axion field typically evolves into a complex network of global strings, domain walls, and oscillons \cite{Hogan:1988mp,Kolb:1993zz}, which must be simulated numerically (e.g. \cite{Buschmann:2019icd}).
On the other hand, if the PQ symmetry breaks before the end of inflation, the average axion density is typically characterized by a single misalignment value within the entire visible Universe, and is more straightforward to estimate. Nevertheless, even in this case the resulting density of axions is driven by its low energy potential, and is thus extremely sensitive to the behavior of QCD at finite temperature \cite{Gross:1980br,Turner:1985si,diCortona:2015ldu,Visinelli:2014twa,Wantz:2009mi}.  Requiring that the relic abundance of axions in such a scenario saturate the cosmologically observed dark matter density occurs for $f_a \simeq 10^{12}$ GeV for an order one initial misalignment angle.

Accessing wider ranges of axion parameter space either requires one to abandon the notion that the axion makes up all of the dark matter, fine-tuning the misalignment angle, or changing the dynamics at early cosmological times.  In this work, we consider a theory in which the strong coupling dynamically evolves at high temperatures, such that QCD initially confines at a high scale, eventually relaxing back to its observed value today \cite{Ipek:2018lhm}.  This modification directly lifts the axion potential, and opens up new regions of parameter space which would be unnatural for a standard cosmology\footnote{While employing different dynamics and operating at a different energy scales, it shares a common theme with Refs.~\cite{Dvali:1995ce,Co:2018phi,Co:2018mho}}.

This article is organized as follows: in Section~\ref{sec:model}, we review the module which allows for dynamical evolution of the QCD coupling during early times, and discuss its implications for the axion potential.  In Section~\ref{sec:abundance}, we compute the resulting abundance of axions under different assumptions concerning in which era the early confinement takes place.  In Section~\ref{sec:results}, we summarize the numerical results of these calculations.  We reserve Section~\ref{sec:outlook} for our conclusions.

\section{Early QCD Confinement}
\label{sec:model}
In this section, we outline a simple module which promotes the strong coupling constant to a dynamical quantity capable of triggering early confinement, and its impact on the mass of the QCD axion during this phase.

\subsection{Dynamical SU(3) Coupling}
Following Ref.~\cite{Ipek:2018lhm} we introduce a scalar field $\phi$ which is a singlet under the SM gauge groups and couples to the gluon field strength via the non-renormalizable operator
\be \label{eq:coupling}
-\frac{1}{4}\left(\frac{1}{g_{s0}^2}+\frac{\phi}{M_\star}\right)G_{\mu\nu}G^{\mu\nu}\,,
\ee
where $g_{s0}$ is the value of the effective strong coupling constant at $\langle\phi\rangle=0$. The ultraviolet scale $M_\star$ represents the typical mass scale of the sector mediating interactions between $\phi$ and the QCD sector, resulting in the effective interaction of Eq.~\eqref{eq:coupling} at low energies $E\ll M_\star$.

\begin{figure*}
    \centering
    \includegraphics[width=0.9\linewidth]{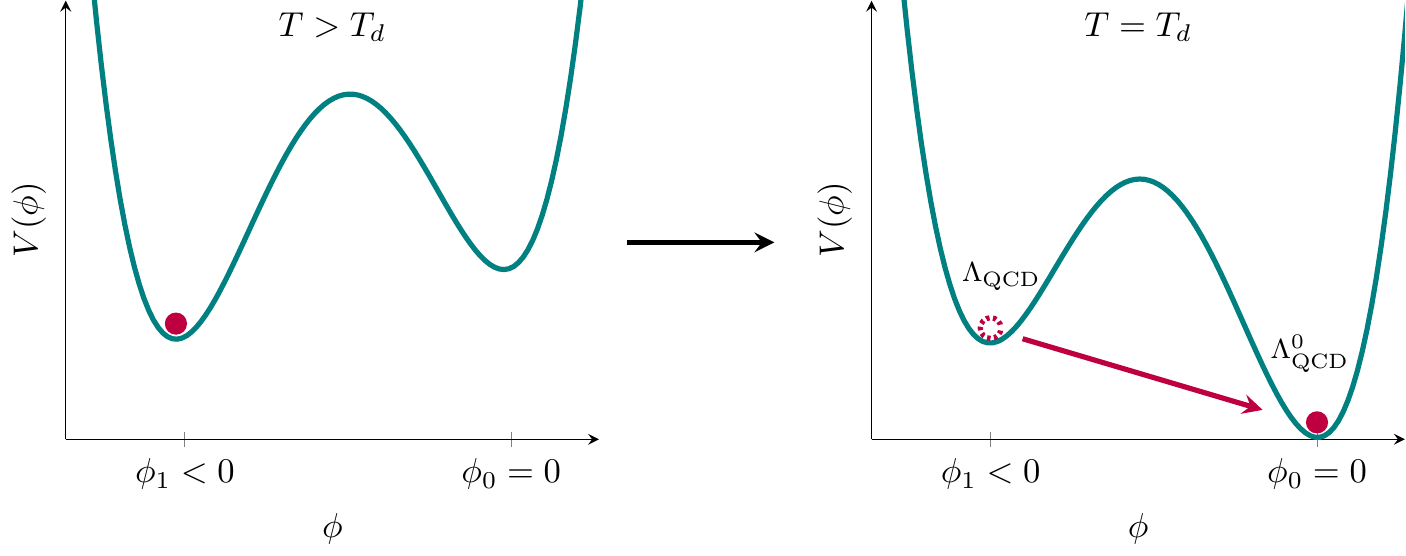}
    \caption{\footnotesize \label{fig:phasetransition} Schematic representation of the potential $V(\phi)$, inducing an effective SU(3) coupling leading to early confinement
    for $T > T_d$, but transitioning to SM-like behavior for lower temperatures.}
\end{figure*}

Due to interactions with particles present in the thermal bath, the vacuum expectation value (VEV) of the scalar field $\phi$ may evolve as a function of the temperature.  If these corrections lead to a larger effective coupling during some early epoch, it may
result in QCD confining earlier than it would under a standard cosmology. 
In principle, the specifics of the interactions between $\phi$ and other fields dictate the shape of its thermal potential and the evolution of $\langle\phi\rangle$ in the early universe. For simplicity, we will assume throughout the paper that the potential is such that $\phi$ tracks an approximately constant value corresponding to a meta-stable vacuum at $\langle\phi\rangle =\phi_1<0$ before transitioning into the true minimum at $\phi_0= 0$  (see Figure~\ref{fig:phasetransition}).  The specific details as to how this is engineered are likely to lead to interesting phenomenology in their own right, but typically less likely to impact the axion abundance through misalignment which is our focus.

In addition to its dependence on $\langle \phi \rangle$, the effective strong coupling runs with the energy scale $\mu$ at one loop:
\be
\frac{1}{\alpha_s(\mu,\vev)}=\frac{33-2n_f}{6\pi}\ln\frac{\mu}{\LambdaQCDSM}+4\pi\frac{\vev}{M_\star}\,,
\ee
where $n_f$ is the number of dynamical quark flavors at the scale $\mu\gtrsim m_f$ and the reference value $\LambdaQCDSM{\simeq 400~\rm MeV}$ corresponds to the QCD confinement scale when $\vev=0$. 
The dependence of the QCD confinement scale on $\vev$ thus reads
\be
\Lambda_{\rm QCD}(\vev)= \LambdaQCDSM \exp\left(\frac{24\pi^2}{2 n_f-33}\frac{\vev}{M_\star}\right)\,.
\ee
Above the electroweak scale (and assuming SM particle content charged under SU(3)), $n_f=6$ and the factor $\frac{24\pi^2}{2 n_f-33}$ is negative.  For $\vev <0$, the value of the confinement scale can be much larger than $\LambdaQCDSM$.  Given these
assumptions,
once the temperature falls below $T_c\sim \Lambda_{\rm QCD}(\phi_1)\gg \LambdaQCDSM$, QCD becomes strong and confines. Later on when $\phi$ transitions to the true minimum at
$\vev = 0$, the confinement scale quickly relaxes to $\LambdaQCDSM$.  
If this happens at $T_d\gtrsim \LambdaQCDSM$, a period of deconfinement can result, until QCD reconfines as in the SM case. 

\subsection{Evolution of the Axion Mass}
\label{sec:axion_mass}
The axion mass depends on the temperature, Higgs VEV, and QCD confinement scale. Analytical and lattice arguments~\cite{Turner:1985si,Wantz:2009mi} suggest the form:
\bea
m_a^2(T)f_a^2 &\approx& \left \{\begin{array}{ll}
\displaystyle m_\pi^2f_\pi^2\bar m, & ~~~(T<\Lambda_\QCD)\,,\\
\displaystyle\zeta m_\pi^2f_\pi^2\bar m\left(\frac{\Lambda_\QCD}{T}\right)^n, & ~~~(T> \Lambda_\QCD)\,,
\end{array}\right.\label{eq:axion_mass}
\eea
where
$\bar m=\sqrt{m_u m_d}/(m_u+m_d)\simeq 0.5$
and the parameters $\zeta$ and $n$ depend on the number of light flavors and thus the
temperature.
For simplicity, we adopt $\zeta=1$ and $n=6.68$ \cite{Wantz:2009mi} throughout.
Note that, the combination $m_\pi^2 f_\pi^2 \simeq m_{\pi0}^2f_{\pi0}^2(v_h/v_h^0)(\LambdaQCD/\LambdaQCDSM)^3$ depends on the confinement scale as well as the ratio between $v_h$ and $v_h^0$, \ie the Higgs VEV at a finite temperature $T$, and its SM value at zero temperature $\sim 246~\rm GeV$.

Early confinement results in two dramatic changes to the usual temperature-dependent Higgs potential of the SM:
\begin{itemize}
\item During early confinement, the thermal bath no longer contains quarks and gluons, but rather the light pseudo-Nambu-Goldstone boson mesons resulting from the spontaneous breaking of the approximate $SU(6)_L\times SU(6)_R$ flavor symmetry to its diagonal subgroup $SU(6)_V$ by the QCD chiral condensate.
\item The quark condensate generates a tadpole term for the Higgs via the SM Yukawa interactions, which shifts the minimum of its potential and thus deforms the usual SM electroweak symmetry breaking (EWSB), and causes it to occur earlier than in the SM, if QCD confines before the SM electroweak phase transition.
\end{itemize}
The resulting thermal Higgs potential is \cite{Ipek:2018lhm,Croon:2019ugf,Berger:2020maa}:
\beq
V(h,T)=\left\{
\begin{array}{lc}
V_0(h)+\displaystyle\frac{T^4}{2\pi^2}\sum_{i=h,W,Z,t}(-1)^Fn_iJ_{B/F}(m_i^2/T^2) & (T>\Lambda_\QCD)\,,  \\
V_0(h)-\sqrt{2}\kappa y_t h+\displaystyle\frac{T^4}{2\pi^2}\sum_{i=h,W,Z,\pi^a}n_iJ_{B/F}(m_i^2/T^2) & (\Lambda_\QCD>T>T_d)\,, \\
V_0(h)+\displaystyle\frac{T^4}{2\pi^2}\sum_{i=h,W,Z,t}(-1)^Fn_iJ_{B/F}(m_i^2/T^2) & (T_d>T)\,,  \\
\end{array}\right. \label{eq:higgs_V_th}
\eeq
where 
\beq
V_0(h)=-\frac{1}{2}\mu^2h^2+\frac{\lambda}{4}h^4\,
\eeq
is the tree level SM Higgs potential,
$n_i$ counts the degeneracy of each particle species, and the bosonic/fermionic thermal functions are
\beq
J_{B/F}(x)=\int_0^{\infty}dy~y^2 \log\left(1-(-1)^F e^{-\sqrt{y^2+x}} \right)\,,\label{eq:JBF}
\eeq
with $F=0/1$ for bosons/fermions.
Notice that, the masses in Eq.~(\ref{eq:higgs_V_th}) are field dependent.
The pion masses are matched to experimental data as in Ref.~\cite{Berger:2020maa}.

\begin{figure}
    \centering
    \includegraphics[width=0.6\textwidth]{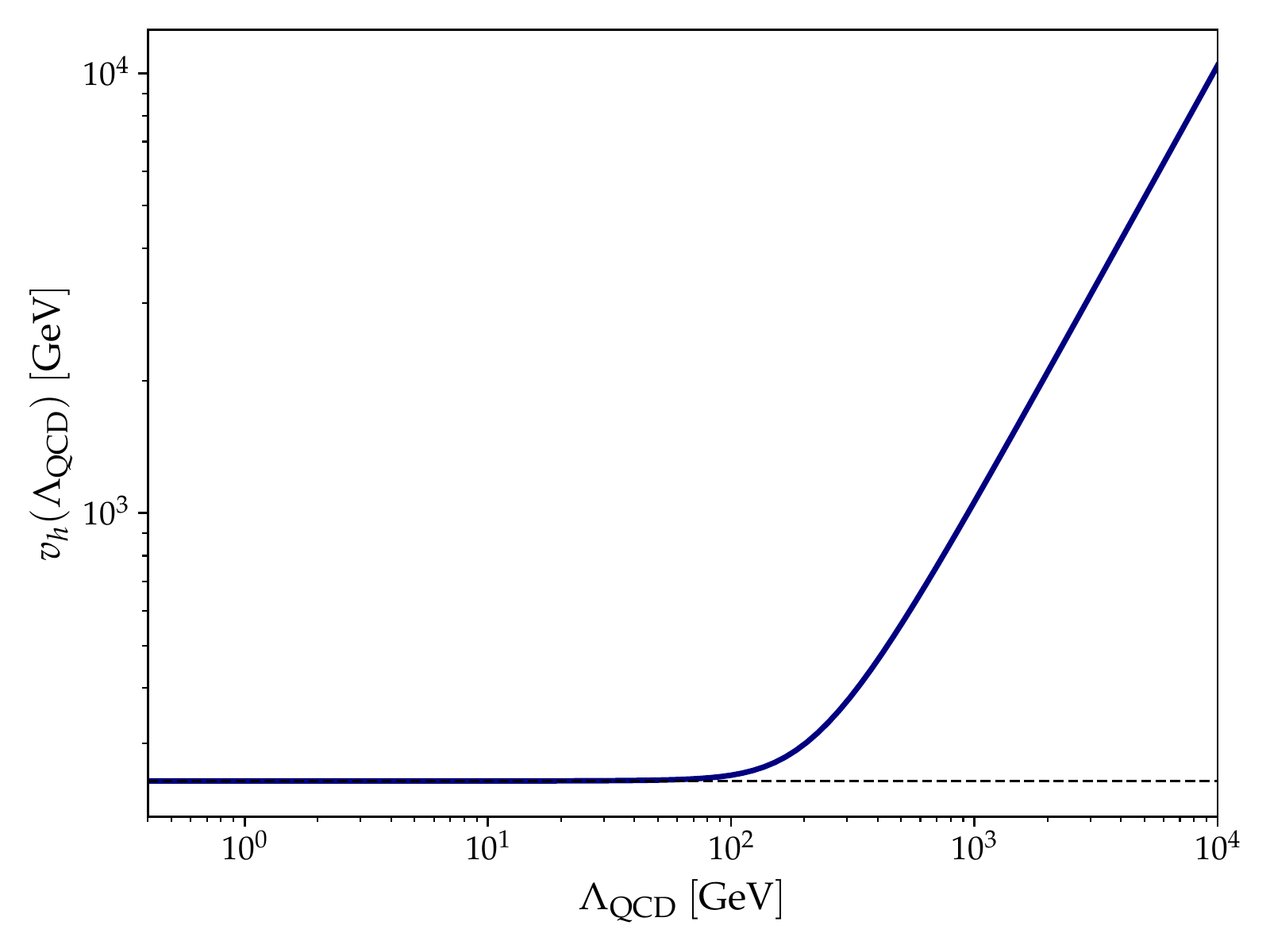}
    \caption{The value of $v_h$ immediately after early confinement is triggered, as a function of the early confinement scale.}
    \label{fig:higgs_vev_LQCD}
\end{figure}

At a given temperature $T$, the Higgs VEV $v_h$ is obtained by minimizing the finite-temperature potential of Eq. (\ref{eq:higgs_V_th}).
The relationship between $v_h$ and $\Lambda_\QCD$ is shown in Figure~\ref{fig:higgs_vev_LQCD}.
In general, if $\LambdaQCD$ is higher than the scale of SM EWSB, $\LambdaEW$, the Higgs VEV is displaced to a significantly higher value than $v_h^0$.
However, if $\LambdaQCD<\LambdaEW$, the thermal corrections are not important and $v_h\sim v_h^0$ for $T<\LambdaQCD$.

In practice, the Higgs VEV does not vary significantly during early confinement \cite{Berger:2020maa}, and it suffices to approximate it as a constant during early confinement.  In fact (as shown below) the final axion abundance only depends on the value of $v_h$ immediately after early confinement is triggered.

\section{Axion Abundance}
\label{sec:abundance}
\subsection{Standard Axion Production}
The relic density of the axions generated via misalignment production can be determined by solving the equation of motion
\beq
\ddot a+3H\dot a + \frac{\partial V_{\rm PQ}}{\partial a}=0\,,\label{eq:eom}
\eeq
whose solution determines the energy density via:
\beq
\rho_a=\frac{1}{2}\dot{a}^2+V_{\rm PQ}.
\eeq
If the oscillation amplitude is sufficiently small, \ie \mbox{$\theta=a/f_a\lesssim \mathcal{O}(1)$}, the potential is approximately quadratic,
and the approximation \mbox{$V_{\rm PQ}\simeq m_a^2a^2/2$} holds, for which the third term in Eq.~\eqref{eq:eom} is simply $m_a(t)^2a$.

The evolution of the axion energy density as governed by Eq.~\eqref{eq:eom} can be heuristically described as follows.
Once the cosine potential develops, the axion field (which is typically displaced away from its minimum) is initially frozen due to Hubble friction while $m_a(T) \ll H(T)$.
In this epoch, the energy density of axions is completely stored as potential energy, which is sensitive to the expansion of the universe only through the evolution of its temperature-dependent mass:
\beq
\rho_a(T)=\frac 1 2 m_a^2(T) ~ a_0^2 \,.\label{eq:rhoa_1}
\eeq
Later as the Universe cools, the axion mass increases and eventually surpasses the Hubble parameter,
at which point the axion field begins to oscillate around its minimum.
The energy density of axions is thereafter described by the energy density contained in the classical oscillations of the pseudo-scalar zero mode,
evolving like matter with a temperature dependent mass:
\beq
\rho_a(T)=\frac{\rho_a(T')}{m_a(T')}\times  \left(\frac{R(T')}{R(T)}\right)^3\times m_a(T)\,,\label{eq:rhoa_2}
\eeq
in which $T'$ is a reference temperature during the oscillation phase, and 
$R$ is the scale factor.  During this phase, $\rho_a/m_a\sim R^{-3}$ behaves like an effective number density.

In practice it is convenient to estimate the relic density by gluing the limiting solutions, Eqs.~(\ref{eq:rhoa_1}) and (\ref{eq:rhoa_2}), together
at the temperature $m_a(T) \sim H(T)$ where the oscillation of the axion field ``just turns on''.
As long as the mass of the axion evolves slowly enough with respect to time, its energy density can be well described by the piecewise equation~\cite{Turner:1983he,Preskill:1982cy,Steinhardt:1983ia,Turner:1985si}:
\beq
\rho_a(T)=
\begin{cases}\displaystyle
\frac{1}{2}a_0^2m_a^2(T)\,,&(T<\Tosc)\\
\displaystyle\frac{1}{2} a_0^2 m_a(T_{\rm osc})m_a(T)\left(\frac{R(T_{\rm osc})}{R(T)}\right)^3\,,&(T>\Tosc)
\end{cases}\label{eq:rhoa_standard}
\eeq
in which $T_{\rm osc}$ is defined through $m_a(T_{\rm osc})=AH(T_{\rm osc})$ with the constant $A$ determined 
phenomenologically by matching Eq.~(\ref{eq:rhoa_standard}) to numerical solutions of Eq.~(\ref{eq:eom}).
In practice, we find that $A \simeq 4$ results in optimal matching (with some dependence on $f_a$).

We refer to Eq.~(\ref{eq:rhoa_standard}) as the ``standard" picture for the evolution of the axion energy density,
and investigate the production of axion DM when the universe undergoes a phase of early QCD (de)confinement as compared to it.

\subsection{Modified Axion Production} 
The dominant effects on axion production in a non-standard cosmological history for which  $\LambdaQCD$ behaves dynamically
can be broadly placed in two classes:
\begin{itemize}
\item \emph{Non-standard evolution of the background cosmology}:
During early confinement, the degrees of freedom in the strong sector transform from free quarks and gluons into bound state hadrons. 
As described above, this can influence the Higgs VEV, and shift the masses of those particles which dominantly receive their mass from it.
Together, these two effects can dramatically modify the effective number of relativistic degrees of freedom in the thermal bath, and thus the expansion history of the Universe.
\item \emph{Non-standard evolution of the axion mass}:
The existence of a larger QCD scale implies that the axion mass is boosted once its mass switches on by the EWSB, 
which could either be triggered by the early confinement itself (if $\LambdaQCD >\LambdaEW$), 
or take place at the usual time (if $\LambdaQCD \leq\LambdaEW$ -- see Eq.~\eqref{eq:axion_mass}).
Such a boost could enable the axion field to start oscillating at an earlier time, resulting in it spending more time in the matter-like
phase and thus suppressing its relic density.
\end{itemize}
We describe each of these classes of effects in more detail below.

\subsubsection{Changes to the Background Cosmology}
The Friedmann equation relates the Hubble parameter with the total energy density of the universe
\beq
H^2=\frac{8\pi G}{3}\rho_{\rm tot}=\frac{\rho_{\rm tot}}{3M_{\rm Pl}^2}\,,
\eeq
where $G$ is the gravitational constant (related to the reduced Planck mass, $M_{\rm Pl} \equiv 1/\sqrt{8\pi G}$)
and $\rho_{\rm tot}$ is the total energy density of the Universe.
Deep within the radiation dominated (RD) epoch, 
$\rho_{\rm tot}$ is well approximated by the energy density of the thermal bath,
\beq
\rho_{\rm tot}(T)\approx \rho_R(T)=\frac{\pi^2}{30}g_{\star}(T) ~ T^4\,,
\eeq
where $g_\star(T)$ is the effective number of relativistic degrees of freedom of the SM thermal bath. 
The resulting Hubble parameter during this epoch can be approximated
\beq
H(T)\approx\sqrt{\frac{\pi^2}{90}}g_\star^{1/2}(T) ~ \frac{T^2}{M_{\rm Pl}}\,. \label{eq:H_T}
\eeq
While the bulk of the effect from the background cosmology
is determined by the timing of the two phase transitions (one to the early confined phase, and the second at de-confinement), there
is some dependence on the detailed evolution of $g_\star$ (and thus, $R$, $T$, and $H$) during each phase transition as well.  As confinement is
an intrinsically non-perturbative process, the physics during the phase transition is typically not amenable to
a perturbative description in terms of a weakly coupled effective field theory.
We describe the physics during the transition using ansatzes which interpolate between limiting behaviors far before and after the
transition, where the description is under control.  We denote by $t_c$ the time at which a transition takes place and $2 \delta t$ its typical duration, 
such that $t_-=t_c-\delta t$ and $t_+=t_c+\delta t$ represent times before the transition begins and after it finishes, respectively.  

During early confinement, the relativistic degrees of freedom in the QCD sector consist of the light pseudoscalar mesons and  
the thermal potential of the Higgs is modified, driving its VEV to increase as described in Section~\ref{sec:model}.
Both of these effects result in a rapid \emph{decrease} in $g_\star$ at the time of early confinement which 
reheats the universe and/or slows down the rate at which its temperature decreases, depending on how fast $g_\star$ is varying.
We model this behavior through the assumption that the scale factor increases, $R_+ > R_-$, whereas the temperature remains approximately constant, 
$T_+=T_- \simeq \LambdaQCD$.  

In contrast, during de-confinement
at $T_d$, there is a sudden \emph{increase} of $g_\star$ either due to the disintegration of hadrons when QCD deconfines (\ie~$T_d>\LambdaQCDSM$), 
or because of the sudden decrease in the pion masses (which scale as $m_\pi^2 \propto \LambdaQCD v_h$), if $T_d<\LambdaQCDSM$.
Thus, during the second transition we treat the scale factor as stagnant, $R_+=R_-$, while the temperature drops: $T_->T_+$.

The conservation of entropy $R_-^3g_\star(T_-)T_-^3=R_+^3g_\star(T_+)T_+^3$ together with Eq.~\eqref{eq:H_T} provide the connection to the
Hubble scale,
\bea
\frac{H_+}{H_-}&=& \left(\frac{R_-}{R_+}\right)^{3/2}  \left(\frac{T_+}{T_-}\right)^{1/2}\,.\label{eq:H_res}
\eea
In the case of an isothermal transition \mbox{($T_+=T_-$)} the scale factor and Hubble scale 
both grow continuously from $R_-$ to $R_+$ and from $H_-$ to $H_+$, respectively, 
though they appear to be discontinuous functions of temperature. 
In the case where the scale factor remains constant ($R_+=R_-$) while the temperature rapidly drops, 
the Hubble scale will decrease continuously as a function of the temperature as
\beq
H(T) = H_-\sqrt{\frac{T}{T_-}}~~~~~~~~(\text{for~} T_-\geq T\geq T_+)\,.\label{eq:H_trans}
\eeq

\subsubsection{Axion Energy Density}
We evaluate the evolution of the energy density of axions following a similar approach to the standard one described above, which treats axion oscillations as immediately
turning on (or off) when the axion mass crosses the Hubble parameter, $m_a(T) = A H(T)$.\footnote{
{It is worth noting that similar ideas, \ie~early oscillation due to the change of mass and the effects on the relic abundance of the corresponding field, have been proposed in previous literature under different contexts \cite{Dienes:2015bka,Takahashi:2015waa,Kawasaki:2015lpf}.}}
We denote the temperature at the crossing points by 
subscripts ``$\uparrow$'' and ``$\downarrow$'' with the former indicating that the mass crosses Hubble from below, and the latter from above.
Moreover, assume that the axion field has not yet begun oscillating before early confinement or EWSB (whichever happens earlier). 

We compare the relic density resulting from early confinement to the standard case by computing their ratio at the moment where the background cosmological
evolution merges and starts tracking a standard cosmology at the completion of de-confinement
(denoted in the notation indicated above by $T_d^+$),
\beq
S \equiv \frac{\rho_a \left( T_d^+ \right)}{\rho_a^{\mathrm{st}} \left( T_d^+ \right)}
\eeq
where $\rho_a^{\mathrm{st}} \left( T_d^+ \right)$ denotes the axion density which would have been obtained for the same parameters under the assumptions of a standard
cosmological history.

As the Universe expands and cools, there are a number of distinct temperature regimes to consider:

\bigskip

\noindent
$(i)$ $T>\Tup$:

As in the standard case, before the axion mass crosses Hubble for the first time, $a$ remains frozen at its initial value $a_0$.
The energy density of axions is stored entirely as potential energy, \ie~$\rho_a(T)\simeq \frac{1}{2}a_0^2m_a(T)^2$ 
which evolves with $m_a^2$ until the axion mass grows comparable to $H$.
Provided the effective boost to the axion mass is large enough,
the first crossing happens when QCD confines (for the first time) or EWSB occurs, whichever is earlier.
For $\LambdaEW\gg\LambdaQCD$, the increase of the axion mass may be more modest, and it may
cross $H$ after EWSB (but still earlier than $\Tosc$ in the standard case).\\

\noindent
$(ii)$ $\Tup > T > T_d^-$:

As the axion mass rises above the Hubble parameter at $\Tup$, the axion field begins to oscillate and behaves like matter (with a varying mass).
At times after the onset of initial oscillations,
\beqn
\rho_a(T)&=&\frac{\rho_a(\Tup )}{m_a(\Tup)}\left(\frac{R(\Tup)}{R(T)}\right)^3 m_a(T)\nn\\
&=&\frac{1}{2} a_0^2 ~m_a(\Tup) m_a(T) \left(\frac{R(\Tup)}{R(T)}\right)^3\,. \label{eq:rhoa_2_EC}
\eeqn
It is important to note that we treat the axion relic abundance as scaling like $\rho_a  \propto R^{-3}m_a$ during all periods in which it is oscillating, which
assumes that the mass itself is changing slowly compared to the classical field oscillations.  This assumption could be violated
during the phase transition in which the effective $\LambdaQCD$ adjusts its value, and strongly depends on the details of the phase transition.
We leave further exploration of these subtleties for future work.

\bigskip
\noindent
$(iii)$ $T_d^->T$:

The subsequent evolution of $\rho_a$ strongly depends on the choice of parameters.
If de-confinement takes place early enough, it can trigger a period where
$m_a$ falls below $H$ again, leading oscillations to damp out.  
However, it may also be the case
that the axion mass never falls below $H$ after oscillations have 
started\footnote{It is also possible that $T_d$ is smaller than $\LambdaQCDSM$ such that hadrons never de-confine.}.

\bigskip

\textit{$(a)$ Damping and Re-oscillation:}

In the case where damping sets in after the initial oscillation phase,
the axion continues to oscillate and its energy density evolves as $R^{-3}$ until $\Tdown$, where its mass falls below the Hubble parameter.
At that point, the oscillations quickly damp out, with $\rho_a$ scaling as potential energy whose evolution is driven by $m_a(T)$.
Therefore, for $T_d^->T>\Tup'$, we have
\beq
\rho_a(T)=\frac 1 2 a_0^2 \frac{m_a(\Tup)m_a^2(T)}{m_a(\Tdown)}\left(\frac{R(\Tup)}{R(\Tdown)}\right)^3\,.
\eeq
At $\Tup'$, the mass once again becomes larger than Hubble, and $a$ begins a new phase of oscillation, evolving as matter.
Therefore, for $T<\Tup'$
\beq
\rho_a(T)=\frac{a_0^2}{2}  \frac{m_a(\Tup)m_a(\Tup')m_a(T)}{m_a(\Tdown)}\left(\frac{R(\Tup)R(\Tup')}{R(\Tdown)R(T)}\right)^3\,.
\eeq
Note that $T_d^+$ could either be larger or smaller than $\Tup'$.
Comparing with Eq.~(\ref{eq:rhoa_standard}) at any temperature $T\leq\min\{\Tup',T_d^+\}$, we obtain 
\beq
S=\frac{m_a(\Tup)m_a(\Tup')}{m_a(\Tdown)m_a^{\rm st}(\Tosc)}\left(\frac{R(\Tup)R(\Tup')}{R(\Tdown)R^{\rm st}(\Tosc)}\right)^3\,,\label{eq:S_a2}
\eeq
in which the superscript ``$\rm st$'' indicates that the corresponding quantity is from the standard 
case.
If $T_d^+>\Tup'$,
the QCD coupling returns to its SM value before oscillation recommences, such that
$\Tup' = \Tosc$ (\ie the axion field remains in a second damped phase at $T_d^+$),
and both $m_a$ and $R$ are both the same as their counterparts in the standard case.
In this situation, Eq.~\eqref{eq:S_a2} then simplifies:
\beq
S=\frac{m_a(\Tup)}{m_a(\Tdown)}\left(\frac{R(\Tup)}{R(\Tdown)}\right)^{3}\,.\label{eq:S_a1}
\eeq

\bigskip

\textit{$(b)$ No Damping:}

In the case where the axion mass never falls below Hubble once oscillations have started,
Eq.~(\ref{eq:rhoa_2_EC}) continues to govern its evolution down to $T_d^+$,
resulting in the ratio, 
\beq
S=\frac{m_a(\Tup)}{m_a^{\rm st}(\Tosc)}\left(\frac{R(\Tup)}{R^{\rm st}(\Tosc)}\right)^{3}\,.\label{eq:S_b}
\eeq
Notice that, since $\Tosc$ is before $T_d^+$, $m_a$ and $R$ are essentially different from their standard case counterparts at $\Tosc$.

\bigskip

Using Eq.~(\ref{eq:H_T}), the ratio $S$ in Eqs. \eqref{eq:S_a2} and \eqref{eq:S_b} can be summarized:
\beq
S\approx
\begin{cases}
\displaystyle\sqrt{\frac{g_\star(\Tdown)g_\star^{\rm st}(\Tosc)}{g_\star(\Tup)g_\star(\Tup')}}\frac{\Tdown\Tosc}{\Tup\Tup'} &~~~~~\text{(a)}\\
\displaystyle\sqrt{\frac{g_\star^{\rm st}(\Tosc)}{g_\star(\Tup)}}\frac{\Tosc}{\Tup} & ~~~~~\text{(b)}
\end{cases}\, .\label{eq:S_cases_new}
\eeq
Note that case (a) simplifies to $\sqrt{\frac{g_\star(\Tdown)}{g_\star(\Tup)}}\frac{\Tdown}{\Tup}$ if $T_d^+>\Tup'$, which is consistent with Eq.~\eqref{eq:S_a1}.
Also, in case (b) if $\LambdaQCD\leq\Tosc$, $\Tup \rightarrow \Tosc$, 
which suggests that if early confinement occurs when the axion field is already oscillating, the relic abundance of the axions is unaffected.
{It is worth noting that the specific value of $\LambdaQCDSM$ has a modest impact on $S$ by modifying the crossing temperatures and pion masses (and thus $g_\star$) during the early confinement phase.
Varying $\LambdaQCDSM$ from 400 MeV to 200 MeV results in an $\mathcal{O}(1)$ factor decrease in $S$ --- in case (a), $S$ changes by less than $10\%$, whereas in case (b), the change is at most $\sim 35\%$ due to a smaller $\Tosc$.}

\section{Results}
\label{sec:results}
\begin{figure}
\centering
\includegraphics[width=0.48\textwidth]{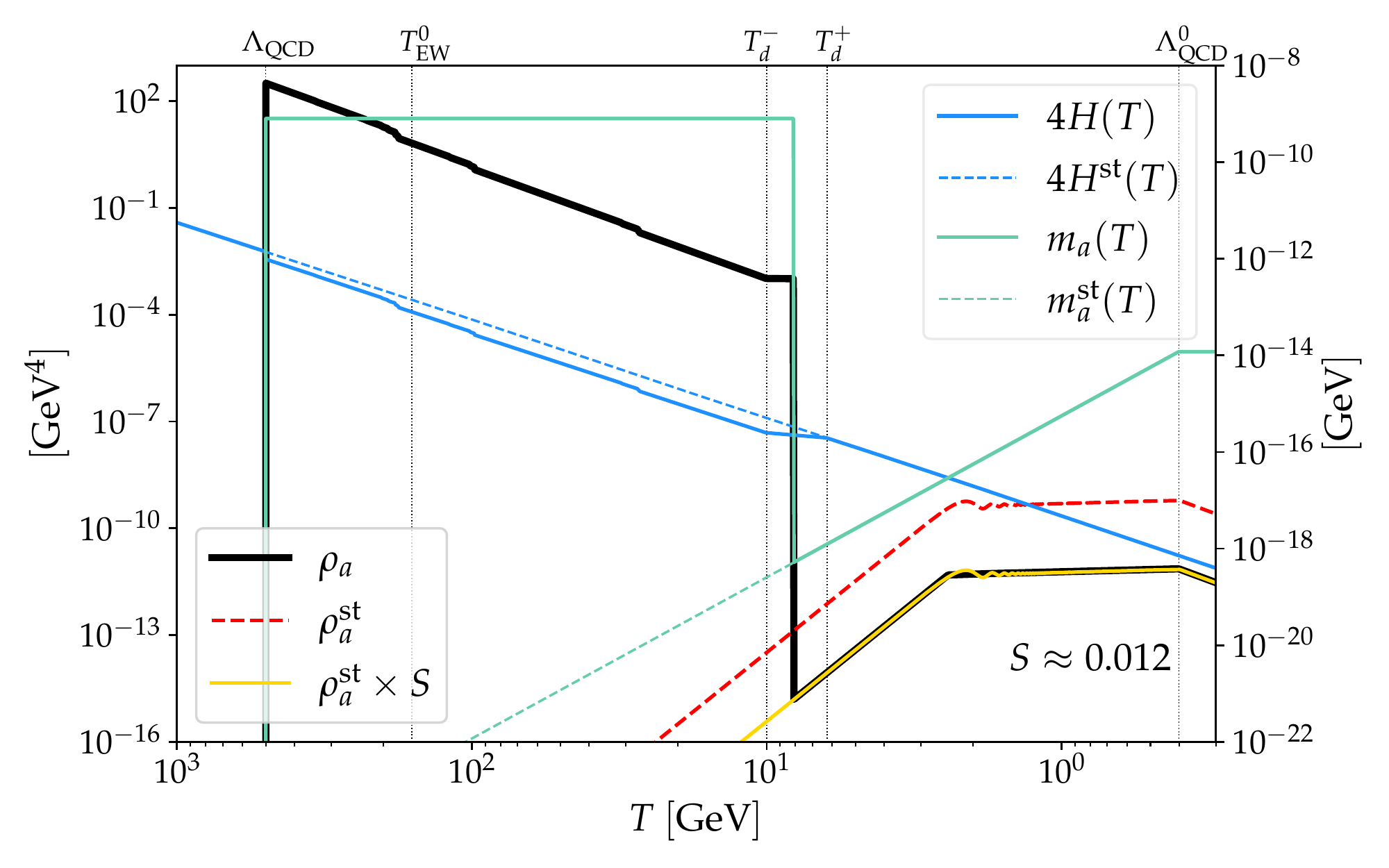}\includegraphics[width=0.48\textwidth]{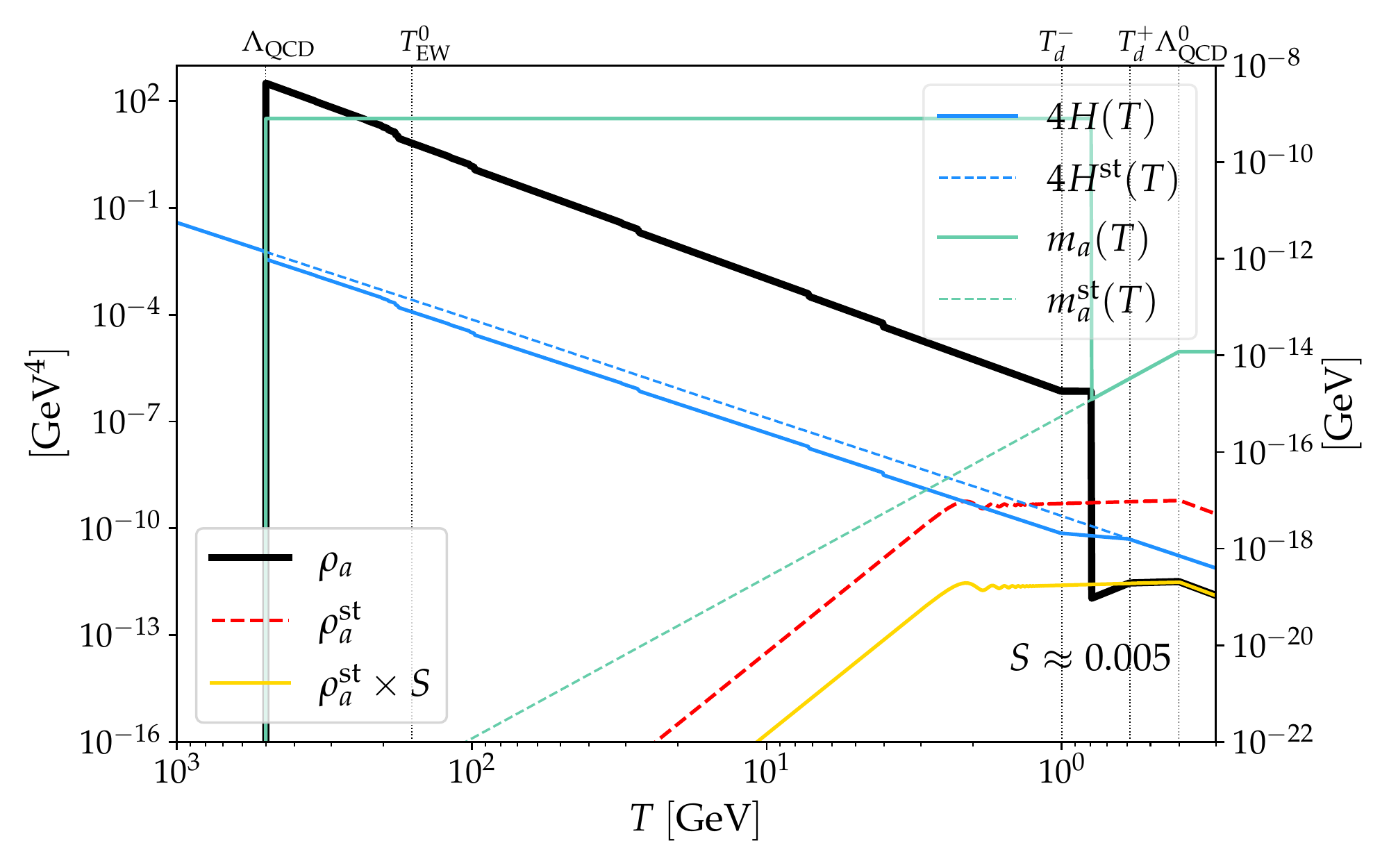}\\
\includegraphics[width=0.48\textwidth]{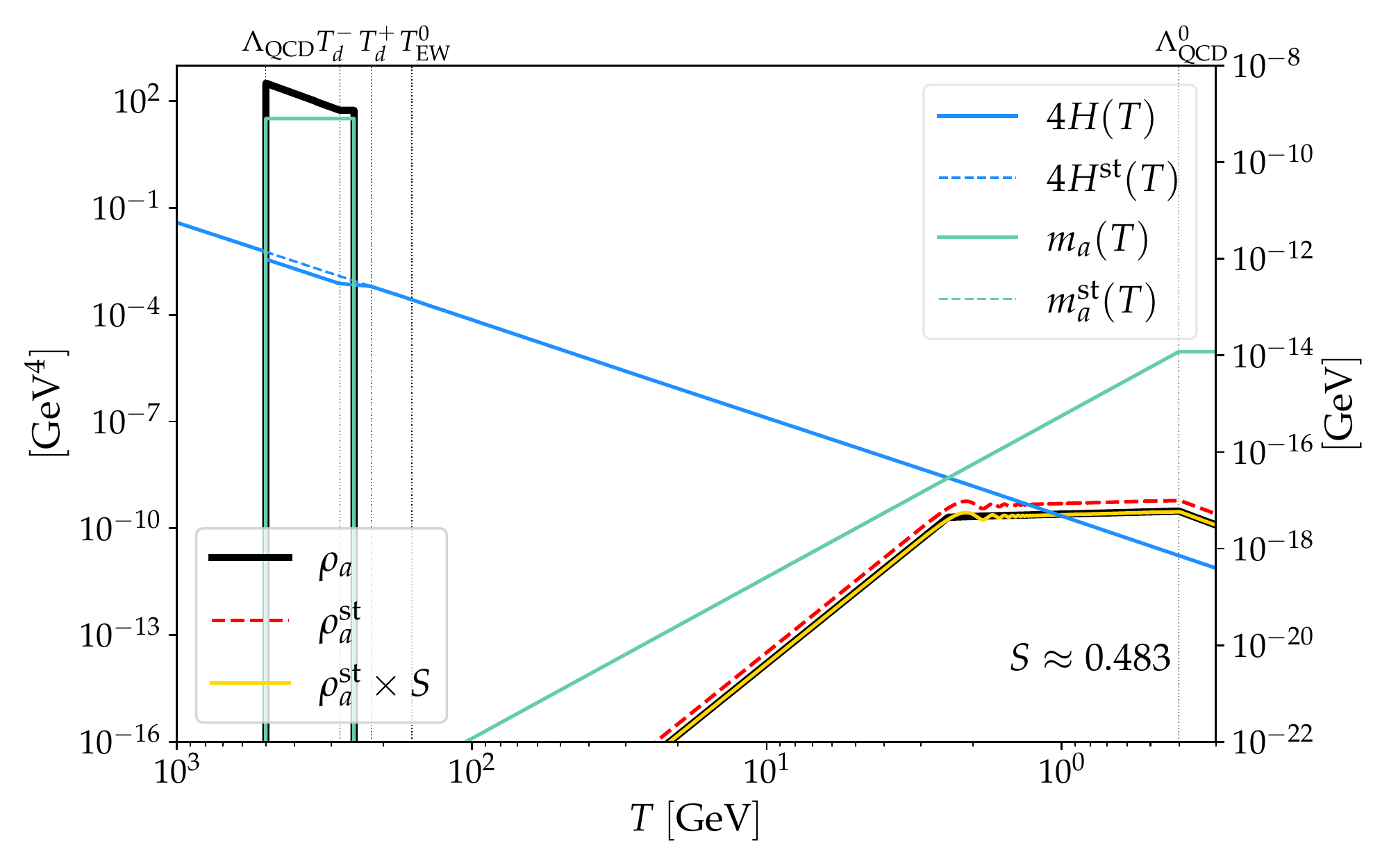}\includegraphics[width=0.48\textwidth]{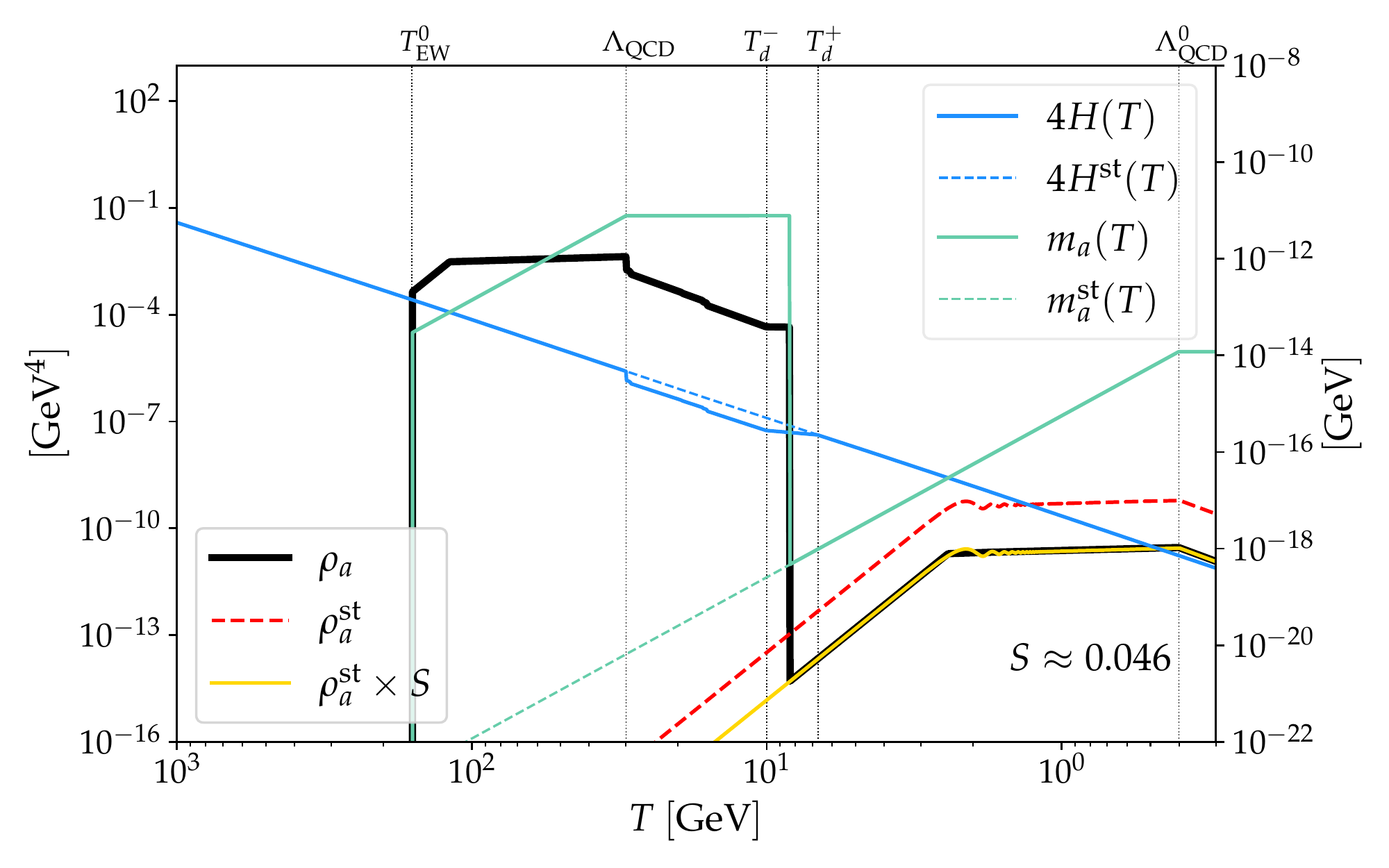}
\caption{Examples of the evolution of $H$, $m_a$, and $\rho_a$ (as indicated) and final {suppression factor} $S$ for four 
choices of parameters:
upper left: $[\LambdaQCD,T_d^-]=[500, 10]~\rm GeV$;
upper right: $[\LambdaQCD,T_d^-]=[500, 1]~\rm GeV$;
lower left: $[\LambdaQCD,T_d^-]=[500, 280]~\rm GeV$; and
lower right: $[\LambdaQCD,T_d^-]=[30, 10]~\rm GeV$.}
\label{fig:evolution_examples}
\end{figure}

Following Eq.~\eqref{eq:S_cases_new}, the axion density resulting in a particular scenario 
depends upon $\Tup$, $\Tdown$ and $\Tup'$, as well as the value of the Hubble parameter at those times. 
We restrict ourselves to cases where initially $\LambdaQCD>\LambdaQCDSM$, and assume the de-confinement transition
is such that $\LambdaQCD$ can be approximated as instantaneously decreasingly to $\LambdaQCDSM$ at $T_d \equiv (T_d^-+T_d^+)/2$.
In practice, we find it convenient to construct interesting scenarios by choosing $\LambdaQCD$ and $T_d^-$,
and computing $T_d^+$ by matching\footnote{For the special case $T_d^->\LambdaQCD$, for which $T_d^+$ 
is undefined, we set $T_d=T_d^-$.}
Eq.~(\ref{eq:H_trans}) with $H^{\rm st}$.  
After obtaining the full evolution of $H$, the values of $\Tup$, $\Tdown$ and $\Tup'$ are computed by solving $m_a(T) = A H(T)$ to determine the time of the corresponding transitions in behavior.

In Figure~\ref{fig:evolution_examples}, we present several examples of cosmological histories which illustrate different cases.  
Each figure shows trajectories based on
the choice of $\LambdaQCD$ and $T_d$ for $H$, $m_a$, and $\rho_a$ (and their standard cosmology analogues) as a function of the (decreasing) temperature scale,
and the final suppression factor $S$ for the axion density compared to the prediction from standard cosmology with $f_a=10^{12}~\rm GeV$ and initial misalignment angle $\theta_0 = 1$.
For the two upper panels and the lower left panel, confinement occurs before $\LambdaEW$ (causing $m_a$ to jump above $H$),
with the difference between the three being where $T_d$ occurs:
$T_d \lesssim \LambdaEW$, causing $m_a$ to fall below $H$ at $T_d$, leading to damping and re-oscillation (upper left);
$T_d \ll  \LambdaEW$, such that $m_a$ never falls below $H$ (upper right);
$T_d>\LambdaEW$, causing two distinct electroweak phase transitions (lower left).  In the lower right panel,
$\LambdaQCD<\LambdaEW$ such that $m_a$ does not immediately cross $H$, but does so at a later time as the temperature decreases.

\begin{figure}
\centering
\includegraphics[width=0.48\textwidth]{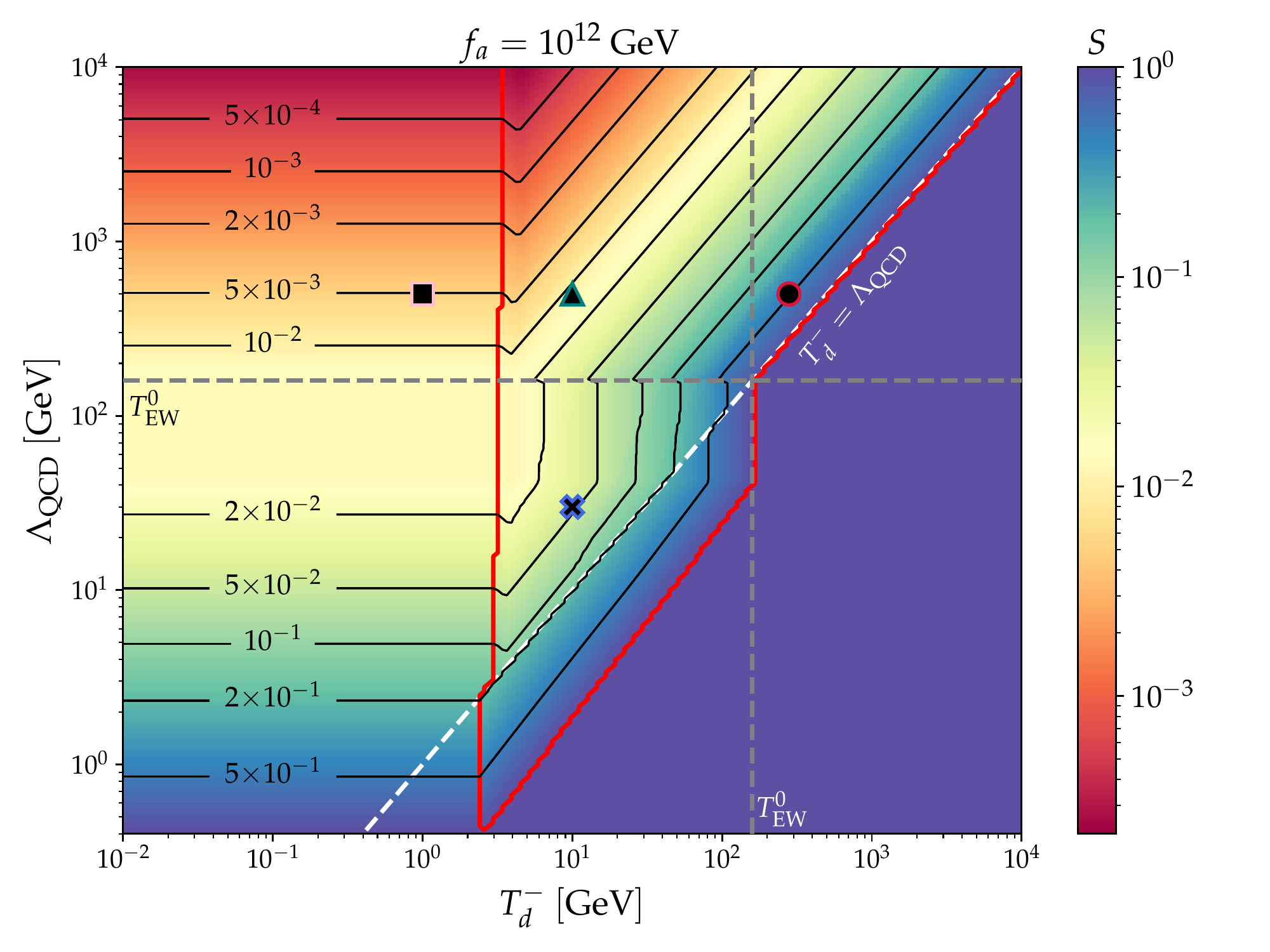}
\includegraphics[width=0.48\textwidth]{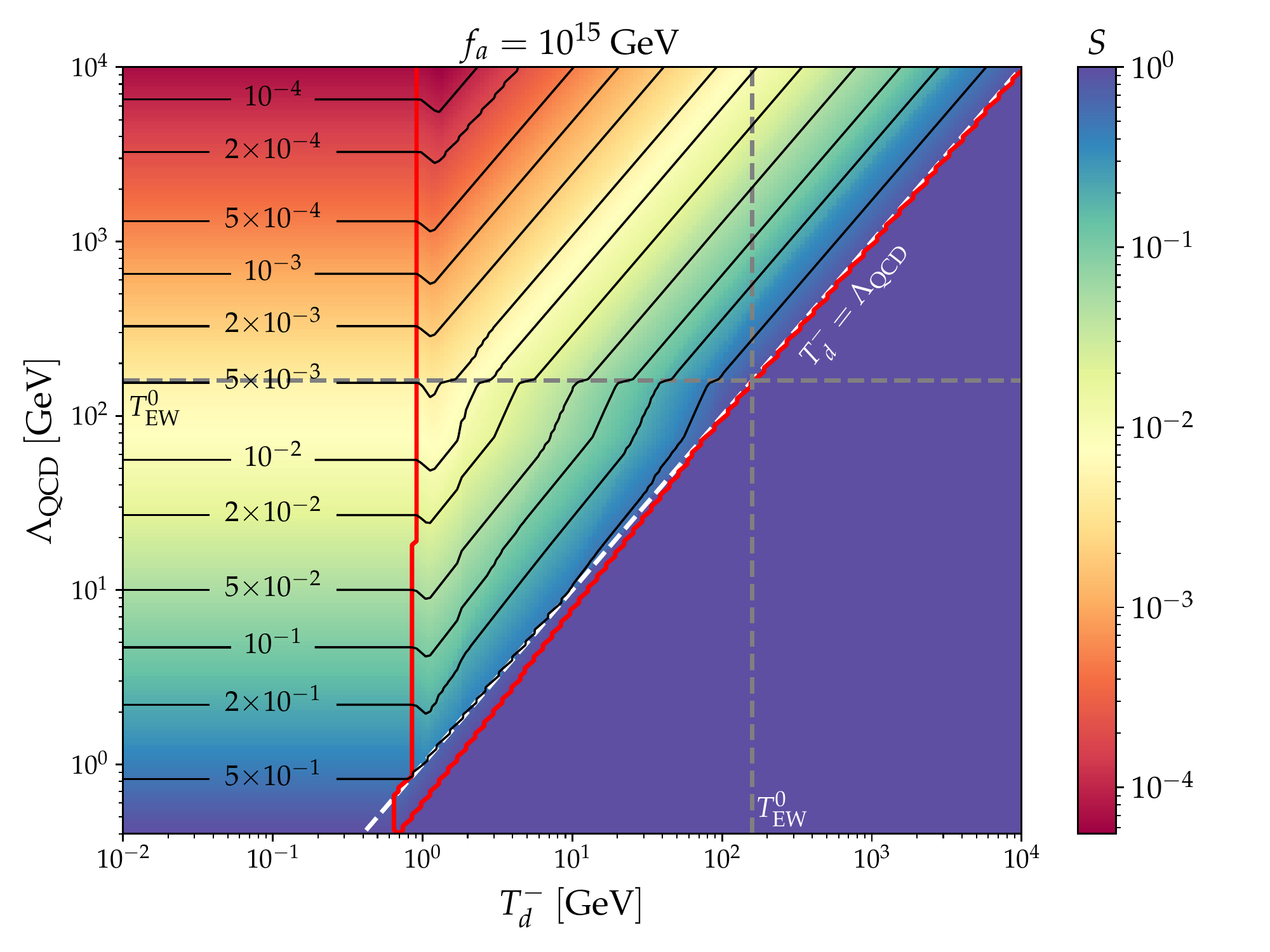}
\caption{Contours of {suppression} factor $S$ in the plane of $T_d^-$ and $\LambdaQCD$ for
$f_a=10^{12}~\rm GeV,~\Tosc\approx 2.4~\rm GeV$ (left) and
$f_a=10^{15}~\rm GeV,~\Tosc\approx 0.67~\rm GeV$ (right).
The red lines bound the region where damping and re-oscillation take place,
the dashed grey lines label the EWSB scale in the SM, and the white dashed line marks the boundary below which early confinement never occurs.
The solid markers in the left panel correspond to the four points shown in Figure~\ref{fig:evolution_examples}.}
\label{fig:S_scan}
\end{figure}

In Figure~\ref{fig:S_scan}, we show contours of $S$ in the plane of $T_d^-$ and $\LambdaQCD$ for two choices of $f_a$.  The example parameter points
from the four panels of Figure~\ref{fig:evolution_examples} are indicated by the triangle (top left), square (top right), circle (bottom left), and cross (bottom right).
The dashed white line indicates $\LambdaQCD=T_d^-$, below which early confinement never occurs.  The red contour delineates the region inside of which
damping and re-oscillation take place.  Broadly, the figure indicates that a period of early confinement can suppress the final axion energy density by orders of magnitude
if early confinement is triggered at scales 
$\LambdaQCD \gg \LambdaQCDSM$.
{Note that a suppression of the axion density can be realized even when $T_d^-\leq \LambdaQCD$ (see the contours below the white dashed line in Figure~\ref{fig:S_scan}).
This is due to the fact that in some parts of the parameter space, having a larger $\LambdaQCD$ still results in early oscillation of the axion field even without an actual phase of early confinement.}

In the damped/re-oscillating region, for $\LambdaQCD > \LambdaEW$ the contours of fixed $S$
are roughly parallel lines since the final density is proportional to $T_d/\LambdaQCD$. 
{In the region where $\LambdaQCD$ is just below $\LambdaEW$, the contours in the left panel appear to be vertical until $\LambdaQCD \sim 40$~GeV --- they are insensitive to the value of $\LambdaQCD$.
In this region, although the axion is massless when $T>\LambdaEW$, as soon as the SM electroweak phase transition occurs, the axion mass jumps above the Hubble parameter such that the axion field starts oscillating immediately at $\LambdaEW$.}
However, as $f_a$ increases, such behavior becomes less obvious, \eg~in the right panel, there is just a slight change of slope when $\LambdaQCD$ becomes smaller than $\LambdaEW$.
As $\LambdaQCD$ further decreases, {the axion mass is less enhanced after $\LambdaEW$. 
The oscillation thus starts later, and again depends on $\LambdaQCD$.}
Outside of the damped region, the axion field never ceases oscillating once it starts, and $S$ is essentially determined 
by $\Tup$, leading to contours which are parallel horizontal lines.

\begin{figure}
\centering
\includegraphics[width=0.6\textwidth]{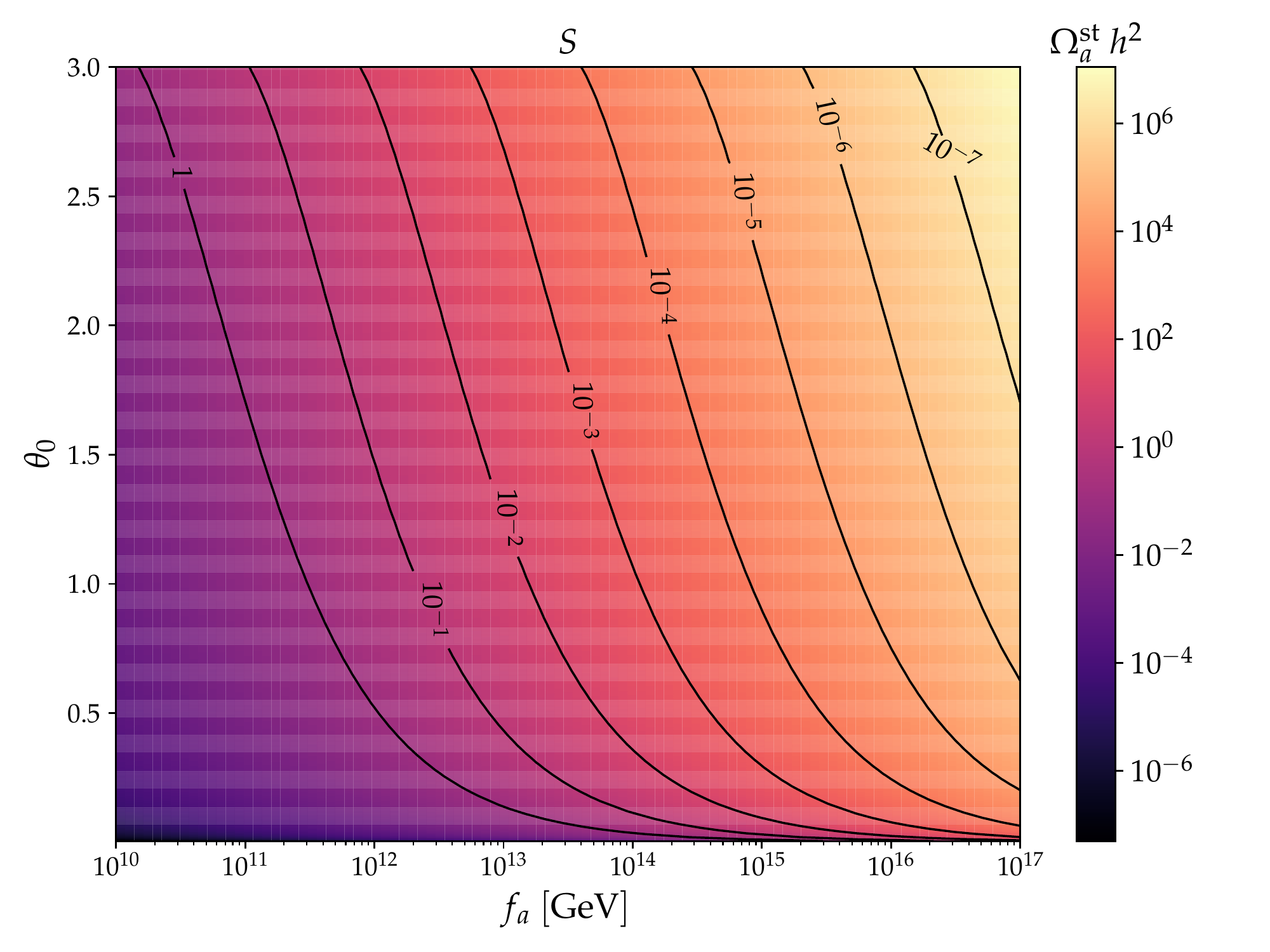}
\caption{Contours in the plane of $f_a$ and the initial misalignment angle $\theta_0$ of the $S$ necessary
to bring the standard cosmology axion relic abundance (indicated by the color map)
into agreement with the observed density of dark matter.}
\label{fig:ftheta}
\end{figure}

Since a dynamical $\LambdaQCD$ can suppress the axion relic abundance, it can be exploited to enlarge the parameter space for QCD axions in regions in which they would otherwise over-close the Universe.
In Figure~\ref{fig:ftheta}, we show the relic abundance\footnote{An anharmonic factor was taken into account for large values of $\theta$ \cite{Marsh:2015xka,Visinelli:2009zm}.} of axions on the plane of $f_a$ and $\theta_0$ in the standard scenario (color map).
In the same plane, contours of $S\leq 1$ indicate the  suppression required to obtain the correct dark-matter relic abundance,
and the region to the left of the $S=1$ contour leads to underproduction of axion dark matter in the standard case.
Taking the reference values $f_a= 10^{12}$ and $f_a=10^{15}$~GeV as examples, the suppression factor needed to obtain the correct relic abundance could be as small as $\sim 10^{-2}$ and $10^{-5}$, respectively.
{From Figure~\ref{fig:S_scan}, we observe that such values of $S$ are easily obtained.
Thus, these regions of parameter space could be resurrected by making $\LambdaQCD$ dynamical.}
In particular, Figure~\ref{fig:ftheta} tells us that regions with a large Peccei-Quinn scale $f_a\gg 10^{12}$ GeV, which corresponds to axions much lighter than in the standard case, can lead to the correct relic abundance in our framework.

\FloatBarrier

\section{Conclusion}
\label{sec:outlook}

The axion is an ideal, well-motivated dark matter candidate whose existence would simultaneously explain the evolution of the Universe and the absence of CP-violation in the
strong nuclear force.  We explore the possibility that the strong nuclear coupling is dynamical, causing the QCD sector to confine much earlier than it would in the Standard Model.
Such a change with respect to the standard cosmology leads to a rich set of possibilities for the evolution of the primordial axion field, which largely depend on the
value of the early confinement scale $\LambdaQCD$ and the temperature at which it evolves to the low energy value observed experimentally today.  We map out the range
of possibilities, which allow the axion to undergo multiple distinct phases of damping and oscillation, 
depending on the interplay between the evolution of its mass and the Hubble parameter throughout the history of the Universe.

We derive analytical estimates of the resulting factor suppressing the final density of axions to the one expected in a standard cosmological history.  We find that a 
theory with a dynamical
QCD confinement scale opens up a wide range of parameter space in which the axion relic density could match the observed dark matter in the Universe
compared to the standard case.
Our results highlight the importance to investigating the possibility of non-standard early cosmologies when considering the range of parameters consistent with
cosmological observations.
{In the light of emerging experiments which can be used to probe axions in the ultra-light mass region, such as IAXO \cite{Armengaud:2014gea}, CASPEr \cite{Budker:2013hfa}, ABRACADABRA \cite{Kahn:2016aff}, KLASH \cite{Alesini:2017ifp}, DANCE \cite{Michimura:2019qxr}, SHAFT \cite{Gramolin:2020ict} and MAGIS \cite{Abe:2021ksx},
our results thus provide extra motivation for experimental efforts along these lines from a theoretical perspective.}

\section*{Acknowledgment}
The authors are grateful for conversations with Luc Darm\'e, and Seyda Ipek.  The work of TMPT was supported in part by
the U.S.~National Science Foundation Grant No.~PHY-1915005.
The work of LH is funded by the UK Science and Technology Facilities Council (STFC)
under grant ST/P001246/1 and was partially supported by the Department of Energy under Grant DE-SC0009913 during part of the realization of this work.
FH is supported by the National Natural Science Foundation of China under Grants No. 12025507, 11690022, 11947302, 12022514 and 11875003; and is supported by the Strategic Priority Research Program and Key Research Program of Frontier Science of the Chinese Academy of Sciences under Grants No. XDB21010200, XDB23010000, and ZDBS-LY-7003.


\bibliographystyle{JHEP}
\bibliography{main}
\end{document}